\documentstyle[amssymb,aps,epsf]{revtex}

\begin{document}

\twocolumn[\hsize\textwidth\columnwidth\hsize\csname @twocolumnfalse\endcsname

\title{Magneto-shear modes and a.c. dissipation 
in a two-dimensional Wigner crystal}
\author{Yuri G. Rubo}
\address{Centro de Investigaci\'on en Energ\'{\i}a, UNAM, \\
Temixco, Morelos 62580, M\'{e}xico}
\author{M. J. Lea}
\address{Department of Physics, Royal Holloway, University of \\
London, Egham, Surrey TW20 0EX, England}
\date{19 February 1998}
\maketitle

\begin{abstract}
The a.c. response of an unpinned and finite 2D Wigner crystal to electric 
fields at an angular
frequency $\omega$ has been calculated in the dissipative limit, $\omega
\tau \ll 1$, where $\tau ^{-1}$ is the scattering rate. For electrons
screened by parallel electrodes, in zero magnetic field the long-wavelength
excitations are a diffusive longitudinal transmission line mode and a
diffusive shear mode. A magnetic field couples these modes together to form
two new magneto-shear modes. The dimensionless coupling parameter $\beta
=2(c_{t}/c_{l})|\sigma _{xy}/\sigma _{xx}|$ where $c_{t}$ ~and $c_{l}$ are
the speeds of transverse and longitudinal sound in the collisionless limit
and $\sigma _{xy}$ and $\sigma _{xx}$ are the tensor components of the
magnetoconductivity. For $\beta \geqslant 1$, both the coupled modes
contribute to the response of 2D electrons in a Corbino disk measurement of
magnetoconductivity. For $\beta \gg 1$, the electron crystal rotates rigidly
in a magnetic field. In general, both the amplitude and phase of the
measured a.c. currents are changed by the shear modulus. In principle, both
the magnetoconductivity and the shear modulus can be measured simultaneously.
\end{abstract}


%\vskip2pc
]

\narrowtext

\section{Introduction}

\label{sec:intro} Free electrons deposited above the surface of liquid
helium form a two-dimensional (2D) conducting system which has been
intensively studied in recent years \cite{review}. The typical surface
concentrations, $n<2\times 10^{9}\;{\rm cm}^{-2}$, for electrons on helium
are small compared to those in semiconductor heterostructures and provide
complementary information about the phases and kinetic properties of
strongly interacting 2D electron systems. It was established long ago \cite
{Grimes,FishHalpPlatz} that at very low temperatures ($T<1\;{\rm K}$) the
electron system undergoes a transition from the liquid to the solid phase,
where the electrons form a Wigner crystal (WC). More recently, there has
been great interest in the kinetics of the WC in an applied magnetic field.
Fascinating non-linear effects have been discovered regarding the WC
magnetoconductivity, including magnetoconductivity hysteresis \cite{Kono95}
and Bragg-Cherenkov scattering \cite{Lea96,BCpaper}.

A specific feature of the 2D electrons on helium is the inability to make
direct ohmic contacts to the electron layer. The kinetic coefficients of 2D
electrons are studied using capacitative coupling between the electrons and
an array of electrodes. The electrons can be held in place above the helium
surface by a positive d.c. voltage applied to the electrodes below the
helium surface. The kinetics of the electron system are studied by applying
a small a.c. voltage, angular frequency $\omega $, to one or more electrodes
and by measuring the a.c. current flow to other electrodes. Due to the
capacitative coupling between the electrodes and the electron sheet, the
electron system behaves as a transmission line, and its kinetic coefficients
define the magnitude and phase of the a.c. currents with respect to the
applied voltage. The connection between the phase shift and the conductivity
is known for the electron liquid \cite{Mehrotra}. In the case when the
electrons form a crystal, however, an additional study is needed because of
the shear modulus, which exists for the electron solid but is absent in the
liquid phase, though the kinematic viscosity may be significant in a highly
correlated liquid.

The goal of this paper is to present such a study and to show the connection
between the microscopic kinetic coefficients of the WC and its macroscopic
a.c. response as observed experimentally. We are not considering the
mechanism of conductivity in unpinned WC \cite{Khazan,Dykman82}, and the
microscopic long-wavelength magnetoconductivity is a parameter in our
theory. It should be noted that the kinetic properties of the WC and the
electron liquid are, in the general case, substantially different. In
particular, the conductivity of the WC on $^{\text{4}}$He is strongly
nonlinear in zero and moderate applied magnetic fields \cite{Kono95,Lea96}
due to the Bragg-Cherenkov mechanism of energy losses \cite{BCpaper}. Strong
magnetic fields, however, smear out the Bragg-Cherenkov resonances and the
conductivity of the WC becomes linear and is the same as in the liquid phase.
\cite{LeaDykman98} Only this case, when the WC kinetics can be described by
linear and local conductivity, will be considered in this paper. However the
shear rigidity of the crystal means that the total response of a finite size
WC to an applied force will be non-local and must be calculated explicitly
for each electrode geometry.

The conductivity of the Wigner solid on normal and superfluid $^{\text{3}}$%
He has recently been measured by Shirahama et al. \cite{3He} and a linear
conduction region is observed, in contrast to the non-linear effects on $^{%
\text{4}}$He. The theory given here should therefore apply to the Wigner
solid on $^{\text{3}}$He in all magnetic fields.

The analysis can be simplified using several limiting conditions. First we
assume that the 2D electrons are in the ''screened limit'', with the
wavelength of the excitations $\lambda \gg d$, the electron-electrode
separation, which acts as a screening length. Secondly, the low frequencies
used experimentally (typically in the audio frequency range) are in the
dissipative limit with $\omega \tau \ll 1$, where $\tau ^{-1}$ is a
scattering rate. We show below that for a Wigner crystal there exist {\it two%
} important low-frequency bulk magnetoplasmon modes. In the case of a small
shear modulus and/or a small applied magnetic field, one of these modes is
similar to the dissipative longitudinal magnetoplasmons in the electron
liquid \cite{Mehrotra}, while the other is connected to a pure shear
diffusive mode. A perpendicular magnetic field mixes these modes, as it does
the two phonon branches of the WC \cite{Bonsall,Fukuyama}. The mixing
parameter behaves as $\mu \left| \sigma _{xy}/\sigma _{xx}\right| $, where $%
\mu$ is the shear modulus, $\sigma _{xx}$ and $\sigma _{xy}$ are diagonal
and off-diagonal components of the conductivity tensor. The mixing becomes
large in classically strong magnetic fields, when $\left| \sigma
_{xy}/\sigma _{xx}\right| \gg 1$, and both modes, which we will refer as
magneto-shear modes, should be taken into account. We then analyze the a.c.
response of the Wigner crystal for the Corbino geometry of electrodes, as
used for magnetoconductivity measurements \cite{Kono95,Lea96,Setup}.

The coupling of the longitudinal and transverse modes in the collisionless
limit, $\omega \tau >>1$, has been analysed and used experimentally to
obtain the shear modulus of the WC by Deville {\it et al.}\cite{Deville84}.
The dynamical matrix for 2D electrons on helium in a magnetic field,
including the coupling to the surface ripplons, was given by Shikin and
Williams \cite{Shikin81}. An analysis of the shear mode resonances of 2D
charged ions \cite{Elliott} below the surface of liquid helium in a magnetic
field has been given by Appleyard {\it et al.} \cite{Appleyard95}. This
system is formally identical to the one studied here, though the parameter
ranges are quite different, and we concentrate here on the dissipative and
screened limit.

The paper is organized in the following way. In Section \ref{sec:plasmons}
we study the magneto-shear modes of a 2D Wigner solid in an applied magnetic
field. In Section \ref{sec:response} we present the theory of the a.c.
response of the Wigner solid in the Corbino geometry. Finally, Section \ref
{sec:discuss} contains the discussion of the results, and conclusions.

\section{Magneto-shear modes}

\label{sec:plasmons} Let the electrons of the WC experience a displacement $%
{\bf u}({\bf r})$, where ${\bf r}=(x,y)$ is the 2D coordinate in the plane
of the crystal. The electric field ${\bf E(r)}$ arising in this plane can be
written as a sum of two terms, ${\bf E(r)=E}_{c}({\bf r})+{\bf E}_{s}({\bf r}%
)$. The first term, ${\bf E}_{c}({\bf r})$, appears due to the change in the
electric-charge density, $\delta \rho ({\bf r)}=en{\bf \nabla }\cdot {\bf u}(%
{\bf r})$ ($n$ is the concentration of electrons), and is the same as for a
normal electron fluid. The second term is specific for the electron solid
and exists due to the shear deformation: 
\begin{equation}
{\bf E}_{s}({\bf r})=\frac{\mu }{ne}{\bf \nabla }\times \left( {\bf \nabla }%
\times {\bf u}({\bf r})\right) .  \label{Eshear}
\end{equation}
If the displacement field is smooth over the distance between the electron
sheet and the electrodes $d$ (typical wave-length $\lambda \gg d$, the
screened limit), one can use the local relation between the change in the
electron density $\delta \rho ({\bf r})$ and the electric field ${\bf E}_{c}(%
{\bf r})$. In this case the main change in the electric potential $\delta V(%
{\bf r})$ is attributed to the image forces in the electrodes, $\delta V(%
{\bf r})=\delta \rho ({\bf r})/C_{s}$, and 
\begin{equation}
{\bf E}_{c}({\bf r})=-\frac{ne}{C_{s}}{\bf \nabla }\left( {\bf \nabla }\cdot 
{\bf u(r)}\right) ,  \label{Ecompr}
\end{equation}
where $C_{s}$ is the capacitance per unit area (for the electrons on helium $%
C_{s}$ is equal to $\varepsilon _{{\rm He}}/(4\pi d)$, with $\varepsilon _{%
{\rm He}}$ being the dielectric permittivity of the liquid helium) \cite
{noteCs}.

In what follows we consider harmonic excitation of the WC, when all relevant
quantities change with time as $\exp (i\omega t)$ (we will not write down
this time dependence explicitly). We assume, as was discussed in Section \ref
{sec:intro}, that the Wigner crystal in the magnetic field $B$ applied
perpendicular to the electron layer exhibits linear and local conductivity
given by a conductivity tensor ${\bf \sigma }\left( \omega \right) $. Than
one can find from Eqs.(\ref{Eshear}),(\ref{Ecompr}) that the density of
electric current ${\bf j(r)}=-i\omega ne{\bf u(r)}$ for the excitations of
the WC satisfies the equation 
\begin{equation}
i\omega C_{s}{\bf j}({\bf r})={\bf \sigma }(\omega )\cdot \left[ {\bf \nabla 
}\left( {\bf \nabla }\cdot {\bf j}({\bf r})\right) -\gamma {\bf \nabla }%
\times \left( {\bf \nabla }\times {\bf j}({\bf r})\right) \right] ,
\label{Main}
\end{equation}
where 
\begin{equation}
\gamma =\mu C_{s}/(ne)^{2}.  \label{gamma}
\end{equation}

Before considering the dissipative magnetoplasmons we note that Eq.(\ref
{Main}) also describes the excitations of the WC in the collisionless limit.
In particular, in the zero-field case ($B=0$), when the components of the
conductivity tensor are $\sigma _{xx}(\omega )=\sigma _{yy}(\omega
)=ne^{2}/(i\omega m)$ and $\sigma _{xy}(\omega )=\sigma _{yx}(\omega )=0$ ($%
m $ is the electron mass), it gives the two phonon branches of the crystal.
In the screened limit these long-wavelength excitations are acoustic in
nature. The velocities of the longitudinal and transverse sound waves are $%
c_{l}=\left( ne^{2}/mC_{s}\right) ^{1/2}$ and $c_{t}=(\mu /mn)^{1/2}$, so
that the dimensionless parameter $\gamma $ entering Eq.(\ref{Main}) is just
their squared ratio, $\gamma =(c_{t}/c_{l})^{2}$.

In the dissipative and screened limit, the longitudinal plasma mode, without
shear, has a wavevector $k$ \cite{usualk}

\begin{equation}
k^{2}=-\frac{i\omega C_{s}}{\sigma _{xx}}.  \label{wvk}
\end{equation}
This diffusive mode is equivalent to a one-dimensional transmission line
mode with distributed capacitance and resistance\cite{ACmode} as in a
coaxial cable at low frequencies, for instance. This is the only relevant
mode in the analysis of the Corbino geometry for a homogeneous, screened 2D
electron fluid \cite{Mehrotra}, with $\mu =0$, and in the WC in zero
magnetic field. In the same limit, $\omega \tau <<1$, in zero field, the
shear mode is also diffusive with a wavevector $q_{s}$

$\label{diffshear}$%
\begin{equation}
q_{s}^{2}=-\frac{i\omega mn}{\mu \tau }=-\frac{i\omega n^{2}e^{2}}{\mu
\sigma _{xx}}\;  \label{qshear}
\end{equation}
but this mode is not excited by the purely radial forces in zero magnetic
field in the Corbino geometry.

We will analyze the dissipative magnetoplasmons using cylindrical in-plane
polar coordinates ${\bf r}\equiv (r,\varphi )$, and we will be interested in
the axisymmetric solutions of Eq.(\ref{Main}) where the current density does
not depend on the angle $\varphi $. In this case, in terms of the radial $%
j_{r}(r)$ and angular $j_{\phi }(r)$ components of ${\bf j}(r)$, Eq.(\ref
{Main}) gives 
\begin{mathletters}
\label{MainAR}
\begin{eqnarray}
i\omega C_{s}j_{r}(r) &=&{\hat{D}}\left[ \sigma _{xx}j_{r}(r)+\gamma \sigma
_{xy}j_{\phi }(r)\right] ,  \label{MainARa} \\
i\omega C_{s}j_{\phi }(r) &=&{\hat{D}}\left[ -\sigma _{xy}j_{r}(r)+\gamma
\sigma _{xx}j_{\phi }(r)\right] ,  \label{MainARb}
\end{eqnarray}
using the diagonal $\sigma _{xx}=\sigma _{yy}$ and off-diagonal $\sigma
_{xy}=-\sigma _{yx}$ components of the conductivity tensor in a magnetic
field.

The operator ${\hat{D}}$, 
\end{mathletters}
\begin{equation}
{\hat{D}=}\frac{d^{2}}{dr^{2}}+\frac{d}{rdr}-\frac{1}{r^{2}}\;,
\label{operD}
\end{equation}
in Eqs.(\ref{MainAR}) is the same as in the Bessel equation, and the
solutions of Eqs.(\ref{MainAR}) are appropriate superpositions of the first
order Bessel and Neumann functions, $J_{1}(qr)$ and $Y_{1}(qr)$. The allowed
values of the wavevector $q$ can be found by substituting $%
j_{r}(r)=AJ_{1}(qr)$ and $j_{\phi }(r)=BJ_{1}(qr)$, which gives the linear
equations 
\begin{mathletters}
\label{eqs4AB}
\begin{eqnarray}
\left( i\omega C_{s}+\sigma _{xx}q^{2}\right) A+\gamma \sigma _{xy}q^{2}B
&=&0,  \label{eqs4ABa} \\
\sigma _{xy}q^{2}A-\left( i\omega C_{s}+\gamma \sigma _{xx}q^{2}\right) B
&=&0,  \label{eqs4ABb}
\end{eqnarray}
for the coefficients $A$ and $B$. It follows from Eqs.(\ref{eqs4AB}) that
there are two magnetoplasmon or magneto-shear wavevectors, $q_{+}$ and $%
q_{-} $, such that 
\end{mathletters}
\begin{equation}
q_{\pm }^{2}=\frac{2k^{2}}{1+\gamma \pm \alpha },  \label{wvectors}
\end{equation}
and the relation between the coefficients is 
\begin{equation}
B^{(\pm )}=-\frac{2}{\sqrt{\alpha ^{2}+\beta ^{2}}\pm \alpha }\frac{\sigma
_{xy}}{\sigma _{xx}}A^{(\pm )}.  \label{BvsA}
\end{equation}
In Eqs. (\ref{wvectors}) and (\ref{BvsA}) we denoted 
\begin{equation}
\beta ^{2}=4\gamma \frac{\sigma _{xy}^{2}}{\sigma _{xx}^{2}}%
\;,\;\;\;\;\;\alpha =\sqrt{(1-\gamma )^{2}-\beta ^{2}}.  \label{alpbet}
\end{equation}

The parameter $\gamma =\left( c_{t}/c_{l}\right) ^{2}$ is small in many
experiments on magnetoconductivity of electrons on helium. Using the
expression for the transverse sound velocity in the WC $c_{t}\simeq
0.5(e^{2}n^{1/2}/m)^{1/2}$ \cite{Bonsall}, we find $\gamma \simeq 2\times
10^{-4}$ for typical electron concentrations $n\sim 10^{8}\;{\rm cm}^{-2}$
and an electron-electrode spacing $d\sim 10^{-2}\;{\rm cm}$. Eqs. (\ref
{wvectors}) and (\ref{alpbet}) show that the mixing of the longitudinal and
transverse dissipative magnetoplasmons is governed by the dimensionless
parameter $\beta =2(c_{t}/c_{l})|\sigma _{xy}/\sigma _{xx}|$. The ratio $%
\sigma _{xy}/\sigma _{xx}$ can be large in classically strong magnetic
fields, so that $\beta $ can become comparable to 1. In this case the
magneto-shear mode wavevectors $q_{+}$ and $q_{-}$ (\ref{wvectors}) are of
the same order of magnitude, both different from the wavevectors $k$, Eq.(%
\ref{wvk}), and $q_{s}$, Eq.(\ref{qshear}). Both magneto-shear modes are
excited in the Corbino geometry and should be taken into account. The
dependences of the magnetoplasmon wavevectors $q_{\pm }$ on $\beta ^{2}$ are
shown in Fig.1. In the $\beta \gg 1$ limit the wavevectors become

\begin{equation}
q_{-}=\sqrt{\frac{\omega \omega _{c}}{c_{t}c_{l}}}\;,\;\;\;\;\;\;\;\;\;%
\;q_{+}=-i\sqrt{\frac{\omega \omega _{c}}{c_{t}c_{l}}}\;,  \label{bigBETA}
\end{equation}
corresponding to propagating and exponentially damped modes. In Eq.(\ref
{bigBETA}) we assumed that the Hall effect is given by the classical
relation $\sigma _{xy}=nec/B$ and $\omega _{c}$ is the cyclotron frequency.

\begin{figure}[htb]
\begin{center}
\epsfxsize=3.2in                %so many inches wide
\leavevmode\epsfbox{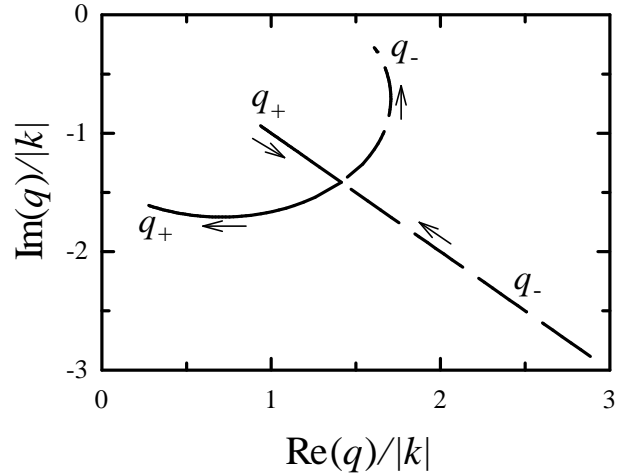}
\end{center}
\caption{The change of the magnetoplasmon wavevectors $q_{+}$ (solid line)
and $q_{-}$ (dashed line) given by Eq.(\ref{wvectors}) with increasing
mixing parameter $\beta ^{2}$, Eq.(\ref{alpbet}), as indicated by arrows.
The wavevectors are presented in units of $\left| k\right| =\left| \omega
C_{s}/\sigma _{xx}\right| ^{1/2}$.}
\end{figure}

\section{Response of the Wigner crystal in the Corbino geometry}

\label{sec:response} In this Section we calculate the spatial distribution
of the electric current density in the Corbino geometry used in Ref.\cite
{Setup,Lea96}. The array of electrodes (Fig.2) consists of a central,
driven, electrode{\rm \ }$A$ with radius $r_{1}$ surrounded by a ring
electrode $E$, which in turn is surrounded by outer, receiving, electrodes $%
B_{1}$, $B_{2}$, and $B_{3}$. Usually the spacing between the $B_{1,2,3}$
electrodes is made to be small, so that one can regard them as one ring
electrode $B$. We denote the outer radius of the $B$ electrode as $r_{2}$,
and the inner radius (which is approximately the same as the outer radius of
the $E$ electrode) as $r_{3}$. Along with a constant voltage applied to the
all electrodes, a small a.c. voltage $V_{0}\exp (i\omega t)$ is applied to
the $A$ electrode. In the case when the sizes of the electrodes are much
greater than the electron-electrode separation, $r_{1,2,3}\gg d$, it
produces an electric field ${\cal E}=({\bf r}/r)V_{0}\delta (r-r_{1})\exp
(i\omega t)$ in the plane of the WC. The induced electric currents can then
be found as the solutions of Eq.(\ref{Main}) with 
\begin{equation}
i\omega C_{s}V_{0}\delta (r-r_{1}){\bf \sigma }(\omega )\cdot \frac{{\bf r}}{%
r}  \label{drvterm}
\end{equation}
added to its right-hand side.

\begin{figure}[htb]
\begin{center}
\epsfxsize=2.0in                %so many inches wide
\leavevmode\epsfbox{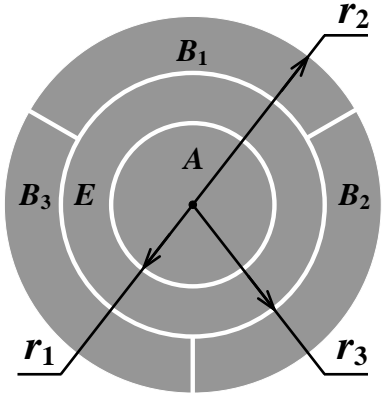}
\end{center}
\caption{The Corbino electrodes geometry. Shown are the driven central
electrode $A$ with radius $r_{1}$, the ring electrode $E$, and the receiving
electrodes $B_{1,2,3}$. Three receiving elecrodes are needed experimentally
to level the system and, since the spacings between them are small, they can
be regarded as one ring electrode with inner radius $r_{3}$ and outer radius 
$r_{2}$.}
\end{figure}

According to the analysis in Section \ref{sec:plasmons}, the radial $%
j_{r}(r) $ and angular $j_{\phi }(r)$ components of the density of electric
current are appropriate superpositions of the cylindrical functions. Since
the solutions cannot be singular at $r=0$, in the region $r<r_{1}$ we have 
\begin{mathletters}
\label{jsm}
\begin{eqnarray}
j_{r}(r) &\equiv
&j_{r}^{<}(r)=A_{1}^{(+)}J_{1}(rq_{+})+A_{1}^{(-)}J_{1}(rq_{-}),
\label{jsma} \\
j_{\phi }(r) &\equiv &j_{\phi
}^{<}(r)=B_{1}^{(+)}J_{1}(rq_{+})+B_{1}^{(-)}J_{1}(rq_{-}),  \label{jsmb}
\end{eqnarray}
while for $r>r_{1}$ both Bessel and Neumann functions are present 
\end{mathletters}
\begin{mathletters}
\label{jgr}
\begin{eqnarray}
j_{r}(r) &\equiv
&j_{r}^{>}(r)=A_{2}^{(+)}J_{1}(rq_{+})+A_{2}^{(-)}J_{1}(rq_{-})  \nonumber \\
&&+A_{3}^{(+)}Y_{1}(rq_{+})+A_{3}^{(-)}Y_{1}(rq_{-}),  \label{jgra} \\
j_{\phi }(r) &\equiv &j_{\phi
}^{>}(r)=B_{2}^{(+)}J_{1}(rq_{+})+B_{2}^{(-)}J_{1}(rq_{-})  \nonumber \\
&&+B_{3}^{(+)}Y_{1}(rq_{+})+B_{3}^{(-)}Y_{1}(rq_{-}).  \label{jgrb}
\end{eqnarray}

The driving term (\ref{drvterm}) produces a jump in the derivative of $%
j_{r}(r)$ at the outer radius of the $A$ electrode, so that we have 
\end{mathletters}
\begin{mathletters}
\label{atr1}
\begin{eqnarray}
j_{r}^{<}(r_{1}) &=&j_{r}^{>}(r_{1})\;,\;\frac{d}{dr}\left(
j_{r}^{<}-j_{r}^{>}\right) _{r=r_{1}}=i\omega C_{s}V_{0};  \label{atr1a} \\
j_{\phi }^{<}(r_{1}) &=&j_{\phi }^{>}(r_{1})\;,\;\frac{d}{dr}\left( j_{\phi
}^{<}-j_{\phi }^{>}\right) _{r=r_{1}}=0.  \label{atr1b}
\end{eqnarray}

The currents are also subject to the following conditions at the outer
boundary of the WC (``stress-free boundary''): 
\end{mathletters}
\begin{equation}
j_{r}^{>}(r_{2})=0\;,\;\frac{d}{dr}\left( \frac{j_{\phi }^{>}}{r}\right)
_{r=r_{2}}=0.  \label{atr2}
\end{equation}

Substituting Eqs.(\ref{jsm}),(\ref{jgr}) into (\ref{atr1}) and making use of
the relations (\ref{BvsA}) between $A_{i}^{(\pm )}$ and $B_{i}^{(\pm )}$ $%
(i=1,2,3)$, one can find 
\begin{equation}
A_{3}^{(\pm )}=\mp \frac{i\pi }{2}\left( \frac{\sqrt{\alpha ^{2}+\beta ^{2}}%
\pm \alpha }{2\alpha }\right) \omega r_{1}C_{s}V_{0}J_{1}(r_{1}q_{\pm }).
\label{coefA3}
\end{equation}
The coefficients $A_{2}^{(\pm )}$ can then be found from Eqs.(\ref{atr2}),
and $A_{1}^{(\pm )}$ from the continuity of currents at $r=r_{1}$. However,
the final expressions are rather cumbersome, and we will present here some
numerical results.

It turns out that the {\it radial} current density $j_{r}(r)$ does not
exhibit strong qualitative changes as the mixing parameter $\beta $ (\ref
{alpbet}) increases, and is similar to the case of the normal electron fluid
where, for pure capacitative coupling with no losses

\begin{mathletters}
\label{jr}
\begin{eqnarray}
j_{r}\left( r\right) &=&i\omega C_{s}V_{0}\frac{\left(
r_{2}^{2}-r_{1}^{2}\right) r}{2r_{2}^{2}},\hspace{0.3in}r\leq r_{1}
\label{jra} \\
j_{r}\left( r\right) &=&i\omega C_{s}V_{0}\frac{r_{1}^{2}\left(
r_{2}^{2}-r^{2}\right) }{2r_{2}^{2}r},\hspace{0.3in}r_{1}<r\leq r_{2}.
\label{jrb}
\end{eqnarray}
It increases (approximately linearly) in the region $r<r_{1}$, and then
decreases monotonously for $r>r_{1}$, going to zero at $r=r_{2}$. More
drastic changes, however, happen to the {\it angular} component of the
density of current $j_{\phi }(r)$, as is illustrated in Fig.3. We note that,
in the case of a normal electron fluid ($\gamma =0$), the angular and radial
components of the density of current are proportional, $j_{\phi
}(r)=-(\sigma _{xy}/\sigma _{xx})j_{r}(r)$ (also seen from Eqs.(\ref{MainAR}%
)). This relation remains approximately true for the WC in a weak applied
magnetic field, when both $\gamma $ and $\beta $ are small. The angular
density of current then follows the dependence $j_{r}(r)$ with a cusp at the
outer boundary of the driven electrode $A$ ($r=r_{1}$). This limiting case
is shown by the thin solid line in Fig.3. Increasing the applied magnetic
field and hence the mixing parameter $\beta $ makes the $j_{\phi }(r)$
dependence much smoother. In the region of classically strong magnetic
fields, when $\left| \sigma _{xy}/\sigma _{xx}\right| \gg 1$ and $\beta \sim
1$, the angular current density increases linearly with radius (thin dashed
line in Fig.3), which corresponds to the {\it rigid rotation} of the Wigner
crystal as a whole in an applied magnetic field. It can be shown that in the
rigid limit, $\beta \gg 1$, the angular density of current is 
\end{mathletters}
\begin{equation}
j_{\phi }(r)=\frac{1}{2}\left( 1-\frac{r_{1}^{2}}{r_{2}^{2}}\right) \frac{%
(kr_{1})^{2}}{1-\frac{1}{24}(kr_{2})^{2}}\frac{\sigma _{xy}V_{0}}{r_{2}^{2}}%
r\;.  \label{jfiinfb}
\end{equation}

\begin{figure}[htb]
\begin{center}
\epsfxsize=3.2in                %so many inches wide
\leavevmode\epsfbox{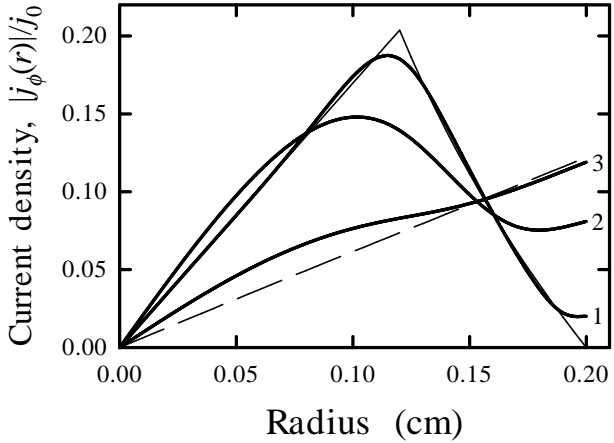}
\end{center}
\caption{The radial dependence of the absolute value of the angular current
density $|j_{\phi }(r)|$ for the mixing parameter $\beta ^{2}=$ 0.01, 0.1,
and 0.5 (curves 1, 2, and 3). The asymptotic dependences for $\beta
\rightarrow 0$ (thin solid line) and $\beta \rightarrow \infty $ (thin
dashed line) are shown. The calculations were carried out for $\left|
kr_{2}\right| =1$, $r_{1}=0.6r_{2}$, and $r_{2}=0.2\;{\rm cm}$. The density
of current is given in units of $j_{0}=(\pi /2)\left| \sigma _{xy}/\sigma
_{xx}\right| \omega r_{1}C_{s}V_{0}$.}
\end{figure}

For the Corbino electrodes shown in Fig.2, the experimentally measured a.c.
current $I$ is collected by the $B$ electrode,
 
\begin{equation}
I=2\pi r_{3}j_{r}(r_{3})  \label{Icurrent}
\end{equation}
assuming that all current which flows through the inner radius $r_{3}$ is
collected. It can be shown that the phase shift between the current $I$ and
the driving voltage $V_{0}$ is $\pi /2$ as $\left| k\right| \rightarrow 0$
(i.e. $\left| \sigma _{xx}\right| \rightarrow \infty $, see Eq.(\ref{wvk})),
so that $I=iI_{0}$ in this limit. The amplitude $I_{0}$ is the same both
with and without a finite shear modulus, 
\begin{equation}
I_{0}=\pi \frac{r_{1}^{2}}{r_{2}^{2}}\left( r_{2}^{2}-r_{3}^{2}\right)
\omega C_{s}V_{0}\;.  \label{I0value}
\end{equation}

For a finite value of $\sigma _{xx}$ the measured current acquires a real
component, in phase with $V_{0}$, and it is convenient to illustrate the
effects of the mixing of the two magneto-shear modes using an Argand diagram
to show the current in the complex $I$-plane. Such Argand diagrams are shown
in Fig.4.

\section{Discussion and Conclusions}

\label{sec:discuss}We have calculated the response of a 2D charged crystal
to the low frequency excitation of a Corbino disk in the screened,
dissipative limit. In zero magnetic field, the electrical response is the
same as for a homogeneous 2D electron fluid, and depends only on the
wavevector $k$ of the transmission line mode. This assumes that the crystal
is not pinned (note that this is in contrast to the WC in semiconductors)
and that no dislocations are generated by the local electron density
gradients, which could give rise to extra losses. For no losses, the
measured current is purely capacitative. For a finite conductivity, the
complex current follows a distinctive locus on an Argand diagram (curve 1 in
Fig.4) as $\sigma _{xx}$ and hence $|k|$ changes. The phase shift away from
pure capacitative coupling, $\theta \propto 1/\sigma _{xx}$ for small phase
shifts.

For a 2D fluid in a magnetic field, the same result is obtained. An
azimuthal Hall current flows $j_{\phi }(r)=-(\sigma _{xy}/\sigma
_{xx})j_{r}(r)\approx -\omega _{c}\tau $ $j_{r}(r)$ but is not detected. The
Lorentz force on the electrons is balanced locally by the a dissipative drag
force $F_{d}(r)\varpropto -v_{H}(r)\varpropto -j_{\phi }(r)$ in classically
strong fields, where $v_{H}(r)$ is the Hall velocity of the electrons. The
measured phase shift $\theta \propto 1/\sigma _{xx}(B)$.

\begin{figure}[htb]
\begin{center}
\epsfxsize=3.2in                %so many inches wide
\leavevmode\epsfbox{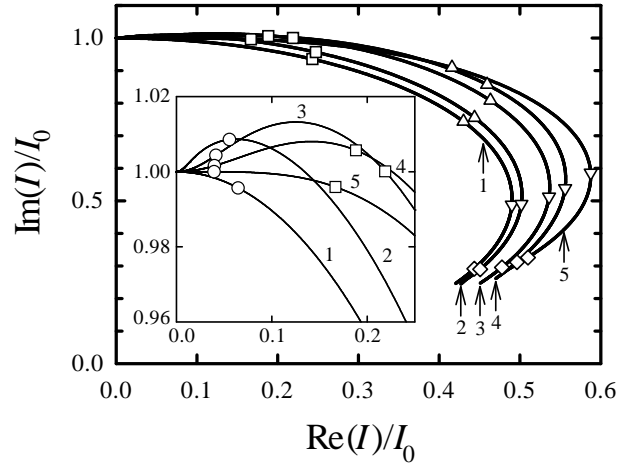}
\end{center}
\caption{The Argand diagram locus for the measured radial current $I$ (see
text) with increasing shear parameter: $\beta ^{2}=$ 0, 0.1, 0.5, 0.99, and
2.0 (curves 1, 2, 3, 4, and 5), calculated for wavevector $\left| k\right|
=\left| \omega C_{s}/\sigma _{xx}\right| ^{1/2}$ values from 1 to 30 cm$%
^{-1} $. For five values of $\left| k\right| $ we have indicated by symbols
the actual positions of the complex vector representing the radial current: $%
\left| k\right| $ = 5 cm$^{-1}$ ($\bigcirc $), 10 cm$^{-1}$ ($\square $), 15
cm$^{-1}$ ($\triangle $), 20 cm$^{-1}$ ($\nabla $) and 25 cm$^{-1}$ ($%
\Diamond $). The upper part of the Argand diagram is shown in enlarged form
in the inset. The components of the current are given in units of $I_{0},$
Eq.(\ref{I0value}). The geometry of the Corbino electrodes was $r_{1}=0.12$
cm, $r_{2}=0.2$ cm, and $r_{3}=0.14$ cm, corresponding to 
Ref.\protect\cite{Lea96,Setup}.}
\end{figure}

In the WC, the shear modulus gives rigidity. In a magnetic field in the
rigid limit, the crystal can only rotate uniformly with $j_{\phi
}(r)\varpropto r$. The Lorentz forces on the capacitative radial currents
produce a torque which is balanced by the torque from the local drag forces $%
F_{d}(r)\varpropto -j_{\phi }(r)$. In this limit, the measured phase shift $%
\theta $ is reduced by a factor $0.54$ as compared to fluid rotation with
the same drag coefficient. The losses, as expressed in the phase shift,
depend on the azimuthal current distribution. Thus the effective Corbino
magnetoconductivity increases by a factor $1.87$ (for small phase shifts) if
the crystal starts to rotate rigidly. At the same time the magnitude of the
current changes and the measured current leaves the normal locus for a 2D
conducting fluid on the Argand diagram, as shown in Fig.4.

In general, the response will lie between the fluid and rigid limits,
controlled by the dimensionless parameter $\beta =2(c_{t}/c_{l})|\sigma
_{xy}/\sigma _{xx}|$ and the propagation of the two magneto-shear modes.
Note that $\beta $ increases as the helium depth $d$ decreases. The
azimuthal current density distribution depends on the value of $\beta $,
being linear in $r$ near the origin and also at the circumference of the
Corbino disc, due to the stress free boundary condition, as shown in Fig.3.
The loci of the measured current $I$, as the conductivity decreases (giving
values of $|k|=|\omega C_{s}/\sigma _{xx}|^{1/2}$ from 1 to 30 cm$^{\text{-1}%
}$), for fixed values of $\beta ^{2}=$ $0$ (fluid), $0.1,$ $0.5,0.99$ and $%
2.0$ are shown in Fig.4. This behavior is a key indication of shear
rigidity. For five values of $\left| k\right| $ we have indicated by symbols
the actual positions of the complex vector representing the radial current: $%
\left| k\right| $ = 5 cm$^{-1}$ ($\bigcirc $), 10 cm$^{-1}$ ($\square $), 15
cm$^{-1}$ ($\triangle $), 20 cm$^{-1}$ ($\nabla $) and 25 cm$^{-1}$ ($%
\Diamond $) for the different values of $\beta $. These points indicate the
change in the current as the shear parameter increases for a given
conductivity. In principle, the measurement of both components of the
complex a.c. current $I$ should enable both the magnetoconductivity $\sigma
_{xx}$ and the shear modulus $\mu $ to be determined, assuming that the Hall
effect is given by the classical result, $\sigma _{xy}=nec/B$, in all cases.

Experimentally, three regions can be approximated in the field dependence of
the magnetoconductivity $\sigma _{xx}$ and hence $\beta $. In low fields in
the fluid phase \cite{Lea97}, the Drude model holds with $\sigma
_{xx}=\sigma _{0}/(1+\omega _{c}^{2}\tau ^{2})$ with a scattering time $\tau 
$ which is independent of magnetic field, due to many-electron effects \cite
{Dykman97}. The Drude model also holds in the WC above liquid $^{\text{3}}$%
He \cite{3He}. This would give $\beta \varpropto B$. But for the WC on
liquid $^{\text{4}}$He, the magnetoconductivity in low fields is non-linear
with $\sigma _{xx}\varpropto 1/B$ with a proportionality constant which
depends on the drive voltage. This would give $\beta $ independent of field,
though the shear-mode theory given here only applies to a linear response.
Above 1 Tesla, the magnetoconductivity in the WC is much less field
dependent \cite{LeaDykman98}, and is almost independent of drive voltage. A
constant $\sigma _{xx}$ gives $\beta \varpropto 1/B$. Hence experimentally $%
\beta $ should increase with magnetic field, pass through a maximum and
decrease again at higher fields. An analysis of recent experimental data in
the WC will be given elsewhere.

In conclusion, we have shown that strong enough applied magnetic field mixes
the diffusive longitudinal plasma and shear modes of 2D Wigner crystal,
forming two diffusive magneto-shear modes. Both coupled modes contribute in
the a.c. response of the Wigner crystal. This gives a crossover from the
liquid-like response of the Wigner crystal at small values of the coupling
parameter $\beta $ to rigid a.c. behavior in the strong coupling limit. Both
the shear modulus and dissipative magnetoconductivity of the 2D Wigner
crystal can, in principle, be measured simultaneously from the amplitude and
phase of the a.c. response.

\section{Acknowledgments}

We are grateful to Mark Dykman for illuminating discussions. One of us
(Y.G.R.) wishes to thank Royal Holloway, University of London for the warm
hospitality extended to him during his visit, when some of this work was
done, partly supported by the Royal Society.

%\newpage

\end{document}